\documentclass[aps,prl,twocolumn,showpacs,a4paper,unsortedaddress,
amsmath,amssymb,byrevtex]{revtex4}

\usepackage[latin1]{inputenc}
\usepackage{graphicx}
\usepackage{dcolumn}
\usepackage{bm}
\usepackage{hyperref}
\usepackage{times}

\begin{document}
\preprint{ffuov/02-01}

\title{Quantum Interference in Single Molecule Electronic Systems}

\author{R. E. Sparks$^1$}
\author{V. M. Garc\'{\i}a-Su\'arez$^{1,2}$}
\author{D. Zs. Manrique$^1$}
\author{C. J. Lambert$^1$}
\affiliation{$^1$ Department of Physics, Lancaster University,
Lancaster, LA1 4YB, U. K.} \affiliation{$^2$ Departamento de
F\'{\i}sica, Universidad de Oviedo \& CINN, 33007 Oviedo, Spain}

\date{\today}

\begin{abstract}
We present a general analytical formula and an ab initio study of quantum interference in multi-branch molecules.
Ab initio calculations are used to investigate quantum interference in a
benzene-1,2-dithiolate (BDT) molecule sandwiched between gold
electrodes and through oligoynes of various lengths. We show that
when a point charge is located in the plane of a BDT molecule and
its position varied, the electrical conductance exhibits a clear
interference effect, whereas when the charge approaches a BDT
molecule along a line normal to the plane of the molecule and
passing through the centre of the phenyl ring, interference
effects are negligible. In the case of olygoynes,  quantum
interference leads to the appearance of a critical energy $E_c$,
at which the electron transmission coefficient $T(E)$ of chains
with even or odd numbers of atoms is independent of length. To
illustrate the underlying physics, we derive a general analytical
formula for electron transport through multi-branch structures and
demonstrate the versatility of the formula by comparing it with
the above ab-initio simulations. We also employ the analytical
formula to investigate the current inside the molecule and
demonstrate that large counter currents can occur within a
ring-like molecule such as BDT, when the point charge is located
in the plane of the molecule. The formula can be used to describe quantum interference and Fano resonances in structures with branches containing arbitrary elastic scattering regions connected to nodal sites.
\end{abstract}

\maketitle

\section{Introduction}

The field of molecular electronics \cite{ME} is a rapidly
expanding research activity, which bridges the gap between physics
and chemistry. Recently there has been much interest in developing
strategies to control the current through a single molecule
\cite{MJ1,MJ2}. Of the various effects that can be exploited,
quantum interference is expected to play a fundamental role in
long phase-coherent molecules \cite{geof}, where multiple
reflections can occur and in molecules made of rings, where
electrons can follow multiple paths between the electrodes
\cite{method1,method2}. The modification of the electronic
properties of such systems has applications such as the
quantum interference effect transistor (QuIET) \cite{QuIET} and
can potentially to be used for implementing data storage \cite{data},
information processing \cite{Infopro} and the development of
molecular switches \cite{switch}.

In this article, we study quantum interference effects in
molecules between metallic leads using a combination of an
analytical model and large-scale ab-initio simulations. We derive
a versatile analytical formula for the electrical conductance of
molecular structures, which captures quantum interference effects in linear and multi-branch molecules. For linear oligoyne molecules or an atomic chain linking two electrodes, we predict that for odd or even-length chains, quantum interference leads to the presence of a critical energy $E_c$, at which
 the electron transmission coefficient becomes independent of length for odd or even numbers of atoms in the chain. The presence of this
 critical energy in more realistic structures is confirmed by performing an ab initio calculation of electron transmission through
 an oligoyne molecular wire connected to gold electrodes.
We also present
results of an ab-initio numerical simulation on an
electrostatically-gated benzene dithiol (BDT) molecule, attached to
gold electrodes, which is an example of a QuIET. In this calculation, gating is achieved
through the presence of a calcium or potassium ion, which induces
quantum interference as the position of the ion and the molecular orientation are varied. We show
that the qualitative features of this interference effect are captured
by the above analytic formula through an appropriate choice of parameters.
Finally, we note that quantum interference in such multi-branch structures leads to the appearance of large internal counter currents, which exceed the external current carried by the electrodes.

\begin{figure*}
\includegraphics[width=2\columnwidth]{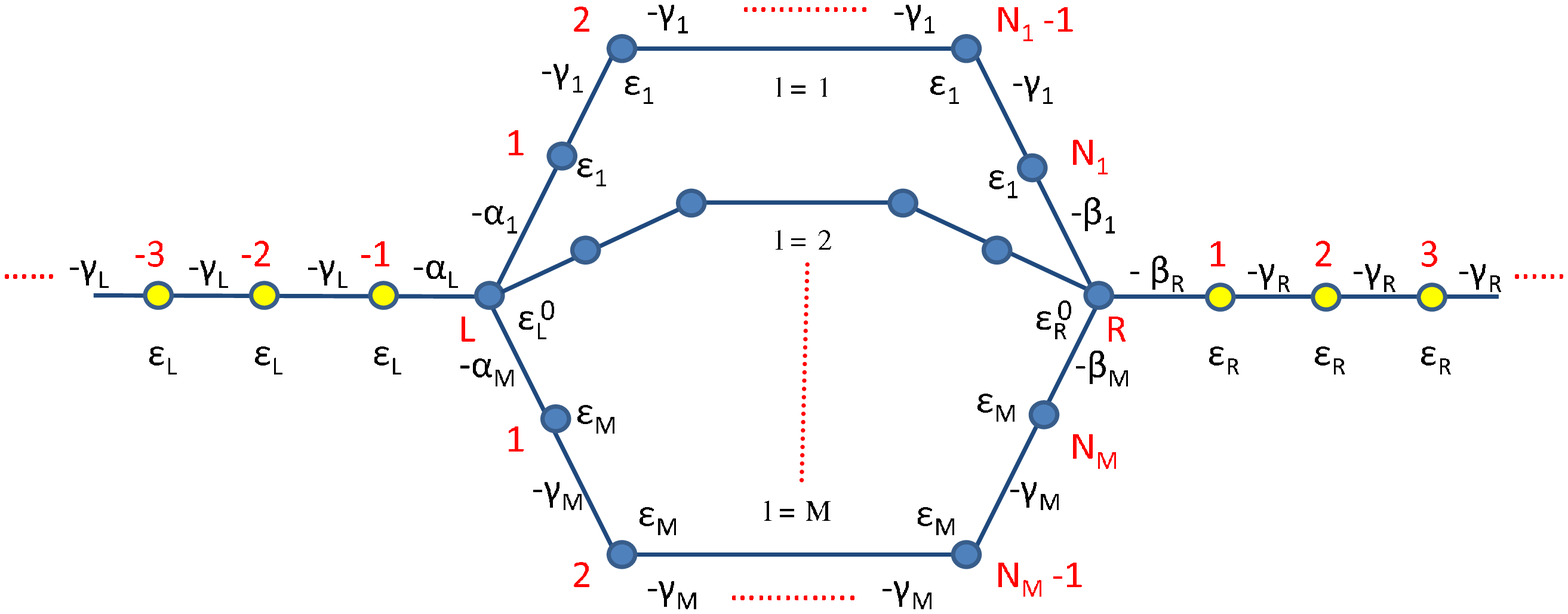}
\caption{\label{config} A multi-branch structure with nodal sites
$L$ and $R$ on the left and right, connected to external
current-carrying leads, by hopping matrix elements $-\alpha_L$ on
the left and $-\beta_R$ on the right and to internals branches
\{$l$\} by hopping matrix elements $-\alpha_l$ and $-\beta_l$
respectively. The energies of the nodal sites are
$\varepsilon_L^0$ and $\varepsilon_R^0$. The site energy and
hopping matrix element of branch $l$ are $\varepsilon_l$ and
$-\gamma_l$ respectively.}
\end{figure*}

\section{An analytical formula for electron transport through  multi-branch structures }

Fig. (\ref{config}) represents a tight-binding (H\"uckel-type)
model of a multi-branch structure, in which each atom is assigned
a single atomic orbital. The structure consists of left and right
leads connected to external electron reservoirs (not shown). The
atoms of the left lead ($L$) are labelled $j= -1, -2, -3, \dots$.
The orbital energy of each atom is denoted $\varepsilon_L$ and
these are coupled to each other by a nearest-neighbour matrix
element $-\gamma_L$. Similarly, the atoms of the right lead
labelled $j= 1, 2, 3, \dots$, are assigned orbital energies
$\varepsilon_R$ and these are coupled to each other by a
nearest-neighbour matrix element $-\gamma_R$. The loop structure
comprises $M$ branches, labelled $l = 1,2, \dots , M$. Branch $l$
possesses $N_l$ atoms, labelled $ n_l = 1,2, \dots , N_l$, with
atomic-orbital energies $\varepsilon_l$, coupled by nearest
neighbour matrix elements $-\gamma_l$. (Note that hopping matrix
elements could be positive or negative and the inclusion of a
minus sign is merely convention. For simplicity, we consider the
case of a real hamiltonian, since in molecules, orbital effects
due to applied magnetic fields are usually negligible.) The
left-most atom ($n_l=1$) of each branch is connected by a matrix
element $-\alpha_l$ to a nodal atom (labelled $L$) of orbital
energy $\varepsilon^0_L$. The latter is connected to the
right-most atom of the left lead by a matrix element $-\alpha_L$.
Similarly, the right-most atom ($n_l=N_l$)of each branch is
connected by a matrix element $-\beta_l$ to a nodal atom (labelled
$R$) of orbital energy $\varepsilon^0_R$, which in turn is
connected to the left-most atom of the right lead by a matrix
element $-\beta_R$.

In the presence of an incoming plane wave from the left, the
solution to the Schr\"{o}dinger equation, $\hat{H}\psi = E \psi$,
in the left lead ($j \le -1$) is of the form

\begin{equation}\label{eq:BL} \psi_j^{(L)} = e^{ik_Lj}+
r(E)e^{-ik_Lj}
\end{equation}

\noindent Similarly, the solution in branch $l$ can be written
\begin{equation}\label{eq:B1}
\psi_{n_l}^{(l)} = A_le^{ik_ln_l}+ B_le^{-ik_ln_l}
\end{equation}

\noindent and the wavefunction in the right lead ($j \ge 1$)  is of
the form

\begin{equation}
\psi_j^{(R)} = t(E)e^{ik_Rj}
\end{equation}

\noindent Finally, the wavefunction on the left and right nodal
atoms will be denoted $\chi_L$ and $\chi_R$ respectively. In the
above equations, $E$ is the energy of the incident electron and
$r(E)$ and $t(E)$ are transmission and reflection amplitudes. For
a given $E$, the dimensionless wavenumbers in the left and right
leads, and in branch $l$ are given by $k_\eta =
\cos^{-1}\left(\frac{\varepsilon_\eta - E}{2 \gamma_\eta}\right)$
where the index $\eta$ is either $L$, $R$ or $l$ respectively. The
corresponding group velocities $(a_\eta/\hbar)dE/dk$ can be
written $(a_\eta/\hbar)v_\eta$, where $a_\eta$ is the atomic
spacing in region $\eta$, $v_\eta = 2\gamma_\eta \sin k_\eta$.
In what follows, we adopt the convention of choosing real values of $k_\eta$,
such that $v_\eta$ is positive and complex values of $k_\eta$, such that Im($k_\eta$) is positive.

Our initial goal is to obtain an expression for the transmission
amplitude $t(E)$, which as shown in the appendix, can be obtained
either by matching wavefunctions at the nodal atoms or by using
Green's functions. According to the Landauer formula, the zero-bias
electrical conductance is simply $(2e^2/h) T(E_F)$, where $E_F$ is
the Fermi energy and

\begin{equation}\label{eq:trans}
T(E)=(v_R/v_L)\vert t(E)\vert^2,
\end{equation}

\noindent which satisfies $T(E) + R(E) =1$, where $R(E)=\vert
r(E)\vert^2$ is the reflection coefficient. In terms of $T(E)$, the  current per unit energy
carried by the left and right leads is $(2e/h)T(E)$ and since
$T(E) \le 1$, the current per unit energy in the left and right
leads cannot exceed $2e/h$. As we shall see below, for $M >1$,
this upper bound does not apply to the current per unit energy
carried by the internal branches, which we denote $(2e/h)I_l$.
Indeed for $M
>1$, $I_l$ can be either positive or negative and is unbounded.

As shown in the appendix, $T(E)$ can be written

\begin{equation}\label{eq:greenT}
T(E) = v_L  \left({\frac{\alpha_L}{\gamma_L}}\right)^2 |G_{RL}|^2
\left({\frac{\beta_R}{\gamma_R}}\right)^2 v_R
\end{equation}

\noindent This expression is very general and shows how the
various contributions combine to control the current through a
single molecule. Equation (\ref{eq:greenT}) shows that the
transmission coefficient $T(E)$ is a product of several factors;
the "group velocities" $v_L$ and $v_R$ describe the ability of the left and right leads
to carry a current, $\left(\frac{\alpha_L}{\gamma_L}\right)$ and
$\left(\frac{\beta_R}{\gamma_R}\right)$ describe the ability of
the couplings between the nodal atoms and the external leads to
transfer electrons and finally $G_{RL}$ describes the ability of a
current from a source at node L to be carried to a current sink at node
R. In this expression, $G_{RL}$ describes propagation from the
nodal site L to at the nodal site R and is sensitive to quantum
interference within the multi-branch structure. Since $v_L$ and
$v_R$ have dimensions of energy, whereas $G_{RL}$ has dimensions
of energy$^{-1}$, the right hand side of equation
(\ref{eq:greenT}) is dimensionless, as expected.

As shown in the appendix, $G_{RL}$ is given by

\begin{equation}\label{eq:green}
G_{RL} = \frac{y}{\Delta},
\end{equation}

\noindent where

\begin{equation}\label{eq:delta}
\Delta = y^2 - (a_L - x_L)(a_R - x_R).
\end{equation}

\noindent In this equation,

\begin{equation}\label{eq:greeny}
y = \sum^{M}_{l=1} y_l,
\end{equation}

\begin{equation}\label{eq:greenxL}
x_L = \sum^M_{l=1} x^L_l
\end{equation}

\noindent and

\begin{equation}\label{eq:greenxR} x_R = \sum^M_{l=1}
x^R_l
\end{equation}

\noindent where

\begin{equation}\label{eq:greenyl} y_l =
\alpha_l\beta_l \sin k_l/[\gamma_l\sin k_l (N_l+1)]
\end{equation}

\begin{equation}\label{eq:greenxlL}
x^L_l =  \alpha_l^2 \sin k_l (N_l)/[\gamma_l\sin k_l (N_l+1)]
\end{equation}

\noindent and
\begin{equation}\label{eq:greenxlR}
x^R_l =  \beta_l^2 \sin k_l (N_l)/[\gamma_l\sin k_l (N_l+1)]
\end{equation}

\noindent Finally, the parameters $a_L$ and $a_R$ are given by
\begin{equation}\label{eq:aL}
a_L = (\varepsilon^0_L-E) - \frac{\alpha_L^2}{\gamma_L}e^{ik_L}
\end{equation}

\noindent and

\begin{equation}\label{eq:aR}
a_R = (\varepsilon^0_R-E) - \frac{\beta_R^2}{\gamma_R}e^{ik_R}
\end{equation}

\noindent Clearly the parameters  $a_L$ and $a_R$ are independent
of the details of the internal branches $l$ and are properties of
the left and right leads and  their respective nodal atoms only.
Properties of the branches are contained within the parameters
$x_L$, $x_R$ and $y$ only. From Eq.(\ref{eq:green}), $T(E)$ will
vanish when y=0. This condition for destructive interference does not depend on the parameters
describing the leads ($\varepsilon_L^0$, $\varepsilon_R^0$,
$\gamma_L$, $\gamma_R$). Nor does it depend on the parameters
describing the contacts to the leads ($\alpha_L$,
$\varepsilon_L$,$\alpha_R$, $\varepsilon_R$). It is a fundamental property of the branches and their couplings to the nodal sites.

As noted in the appendix, equation (\ref{eq:greenT}) is extremely general. With a slight modification of the nodal energies $\varepsilon_L^0$ and $\varepsilon_R^0$, it can be used to describe the effect of Fano resonances due to dangling bonds at the nodes. Furthermore, with a slight redefinition of $y_l$, $x_l^R$ and $x_l^L$, it describes electron transmission arising when the branches are replaced by arbitrary elastic scatterers connected by single bonds to the nodal sites.

An alternative form of equation (\ref{eq:greenT}) is obtained by
writing $\Delta = \Delta_1 + i \Delta_2$, $a_L= \tilde a_L -
i\tilde\Gamma_L$ and $a_R= \tilde a_R - i\tilde\Gamma_R$, where
$\tilde a_L=\varepsilon^0_L-E -(\alpha_L^2/\gamma_L)\cos k_L$ and
$\tilde\Gamma_L=(\alpha_L^2/\gamma_L)\sin k_L$ and similarly for
$\tilde a_R$ and $\tilde\Gamma_R$. With this notation,

\begin{equation}\label{eq:dl}
\Delta_1=y^2-(x_L-\tilde a_L)(x_R- \tilde a_R) +\tilde\Gamma_L\tilde\Gamma_R
\end{equation}
\noindent and
\begin{equation}\label{eq:d2}
\Delta_2=\tilde\Gamma_R(x_L-\tilde a_L) +\tilde\Gamma_L(x_R-\tilde
a_R)
\end{equation}

\noindent and

\begin{equation}\label{eq:TE}
T(E)=\frac{4\tilde\Gamma_L \tilde\Gamma_R y^2}{\Delta_1^2 +
\Delta_2^2}
\end{equation}

Equation (\ref{eq:greenT}) describes the transmission coefficient
of the combined structure and allows us to evaluate the current
per unit energy $(2e/h)T(E)$ due to incident electrons from the
left lead with energies $E$. We shall also be interested in the
current per unit energy $(2e/h)I_l$ carried by branch $l$. As
shown in the appendix, this is given by

\begin{equation}\label{eq:il}
I_l=T(E)y_l/y,
\end{equation}
which clearly satisfies
\begin{equation}\label{eq:isum}
\sum_{l=1}^M I_l=T(E).
\end{equation}

\noindent Unlike $T(E)$, which satisfies $0 \le T(E) \le 1$, $I_l$
can have arbitrary sign and arbitrary magnitude.

\begin{figure}
\includegraphics[width=0.9\columnwidth]{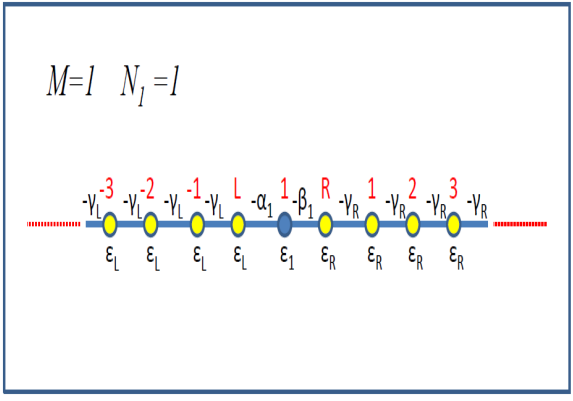}
\caption{\label{Ex1} A single impurity in a 1-d chain}
\end{figure}

Before using equation (\ref{eq:greenT}) to describe quantum
interference within linear and multi-branch molecules, we consider
the simplest choice of a single impurity level, weakly coupled to
 external left and right leads, by matrix elements
$\alpha_1$ and $\beta_1$ respectively, is shown in Fig(\ref{Ex1}).

\noindent This corresponds to the choice $M=1$, $N_1=1$,
$\gamma_L=\alpha_L$ $\gamma_R=\beta_R$,
$\varepsilon_L=\varepsilon^0_L$, $\varepsilon_R=\varepsilon^0_R$.
In this case equation (\ref{eq:greenT}) reduces to the well-known
Breit-Wigner formula

\begin{equation}\label{eq:BW}
T(E) =
{\frac{4\Gamma_L\Gamma_R}{(E-\varepsilon_1+\sigma_L+\sigma_R)^2 +
(\Gamma_L + \Gamma_R)^2}}
\end{equation}

\noindent where
$\sigma_L=(\alpha_1^2/\gamma_L)\cos k_L$,
$\sigma_R=(\beta_1^2/\gamma_R)\cos k_R$,
$\Gamma_L=(\alpha_1^2/\gamma_L)\sin k_L$ and
$\Gamma_R=(\beta_1^2/\gamma_R)\sin k_R$.

\section{Quantum interference in linear molecules or atomic chains.}

\begin{figure}
\includegraphics[width=0.9\columnwidth]{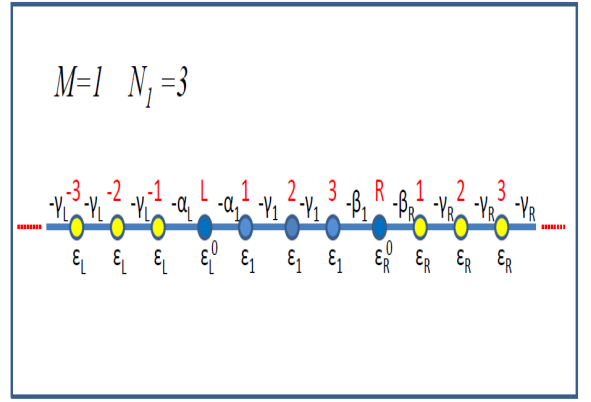}
\caption{\label{Ex2} A schematic of the nodal atoms connected by an atomic chain.}
\end{figure}

\noindent The choice $M=1$ $N_1 >1$ corresponds to the case of
 external left and right leads, coupled by
matrix elements $\alpha_L$ and $\beta_R$ respectively, to nodal
sites $L$ and $R$, which in turn are connected by matrix elements
$\alpha_1$ and $\beta_1$ to an atomic bridge of $N_1$ atoms.
The case $N_1=3$ is shown in Fig(\ref{Ex2}).

For $M=1$, one obtains

\begin{equation}\label{eq:y1}
y=y_1=\frac{\alpha_1\beta_1}{\gamma_1}\frac{\sin k_1}{\sin k_1 (N_1+1)}
\end{equation}

\begin{equation}\label{eq:xL}
x_L=\frac{\alpha_1^2}{\gamma_1}\frac{\sin k_1N_1}{\sin k_1 (N_1+1)}
\end{equation}

\begin{equation}\label{eq:xR}
x_R=\frac{\beta_1^2}{\gamma_1}\frac{\sin k_1N_1}{\sin k_1 (N_1+1)}.
\end{equation}

In the case of a metallic or ``$\pi$ bridge'', $k_1$ will be real.
In the case of a "$\sigma$ bridge", (which acts as a tunnel
barrier), $k_1$ will be imaginary and equation (\ref{eq:greenT})
(or equivalently equation (\ref{eq:TE})) describes electron
transport via superexchange. Equations (\ref{eq:y1}),
(\ref{eq:xL}), (\ref{eq:xR}) highlight a curious feature, which
occurs at a special energy $E_c$, which corresponds to electrons
propagating at the band centre of a $\pi$ bridge and at which
$k_1=\pi /2$. At this energy, $y$, $x_L$ and $x_R$ become
independent of the length $N_1$ of the bridge. On the one
hand, if the bridge contains an even number of atoms (ie if $N_1$
is even), then $x_R=x_L=0$,
 $y=(\alpha_1\beta_1/\gamma_1)(-1)^{(N_1/2)}$ and
\begin{equation}\label{bridgeeven}
T(E_c) =
{\frac{4\tilde\Gamma_L\tilde\Gamma_R(\alpha_1\beta_1/\gamma_1)^2}{((\alpha_1\beta_1/\gamma_1)^2-\tilde
a_L\tilde a_R +\tilde\Gamma_L\tilde\Gamma_R)^2 + (\tilde
a_L\tilde\Gamma_R +\tilde a_R\tilde\Gamma_L)^2}},
\end{equation}

On the other hand, if the bridge contains an odd number of atoms, then $x_L, x_R$ and $y$ diverge and
\begin{equation}\label{bridgeodd}
T(E_c) = {\frac{4\tilde\Gamma_L\tilde\Gamma_R}{(\alpha_1^2\tilde
a_R +\beta_1^2\tilde a_L)^2 + (\alpha_1^2\tilde\Gamma_R
+\beta_1^2\tilde\Gamma_L)^2}},
\end{equation}
\noindent which is independent of the length $N_1$ of the bridge.
This situation can arise, for example, in the case of oligoynes connected to external electrodes.

\begin{figure}
\includegraphics[angle=0,width=0.9\columnwidth]{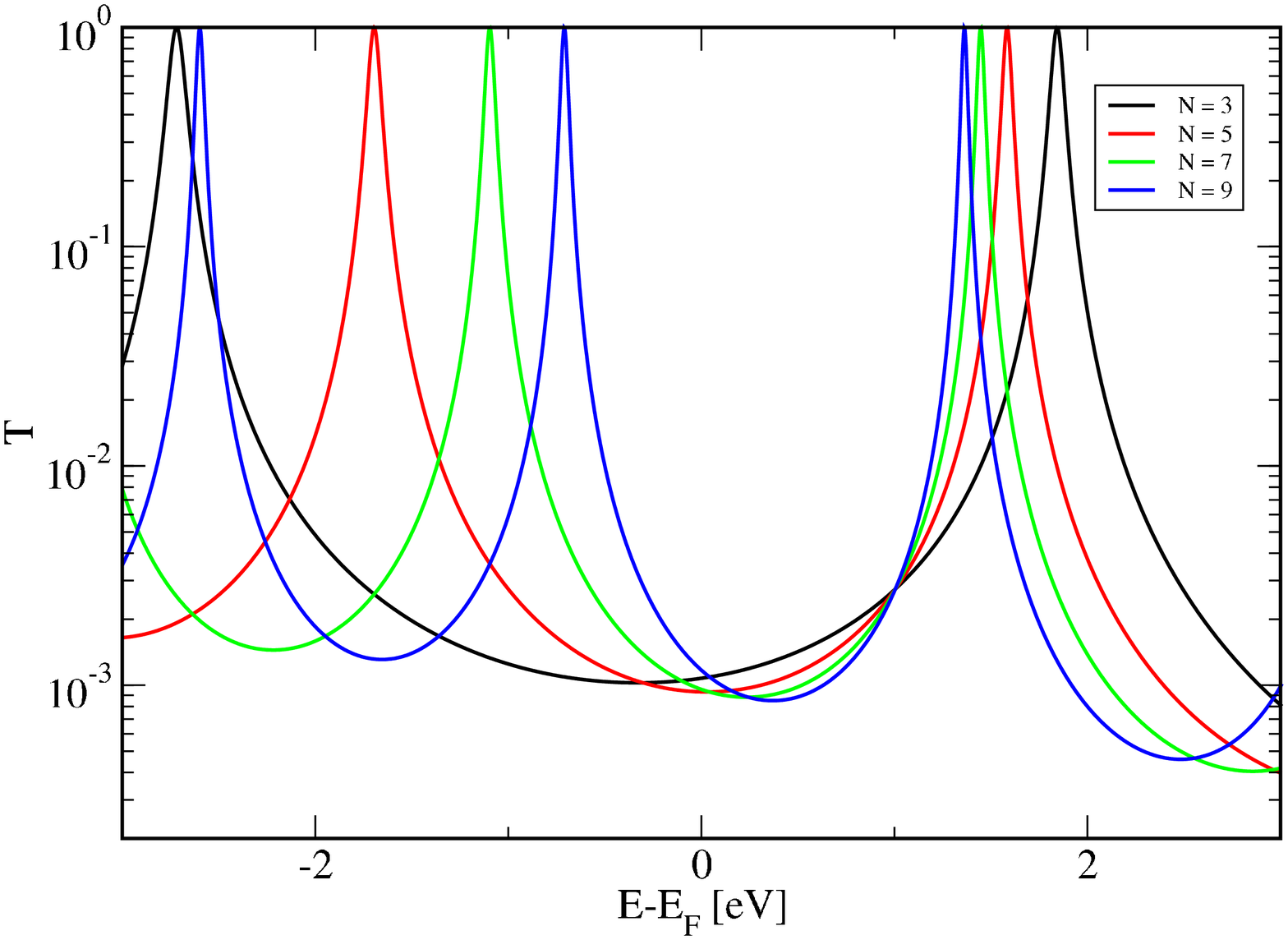}
\caption{\label{olicalc}Transmission functions for increasing
lengths of molecular wire using the general formula. The system is
modelled using the following parameters, in the leads
$\varepsilon_L=\varepsilon_R = 0.0$ and $\gamma_L=\gamma_R=5.0$,
for the contacts $\varepsilon_L^0 = \varepsilon_R^0 = -0.4$, in
the chain $\varepsilon_1 = 0.5 $ and $\gamma_1 =6.0$ and the
coupling between the wire and the electrodes are defined by
$\alpha_L=\beta_R=3.0$. All curves intersect at $E_c \approx 0.5
{\rm eV}$. Close inspection reveals that $E \approx -0.2 {\rm eV}$
the curves approach each other, but do not intersect at a single
energy.}
\end{figure}

These predictions are shown in Fig.(\ref{olicalc}) for increasing
numbers of atoms in the wire, $N_1 = 2,4,6$ and $8$. At the
critical energy $E_c \approx 0.5 {\rm eV}$, all curves intersect.
Consequently, for energies $E$ slightly greater than $E_c$, $T(E)$
will either increases monotonically as the length of the bridge
increases by 2, and for $E$ slightly less than $E_c$, $T(E)$ will
decrease when the length of the bridge increases. This effect is a
clear manifestation of phase-coherent quantum transport.

\begin{figure}
\includegraphics[angle=0,width=0.9\columnwidth]{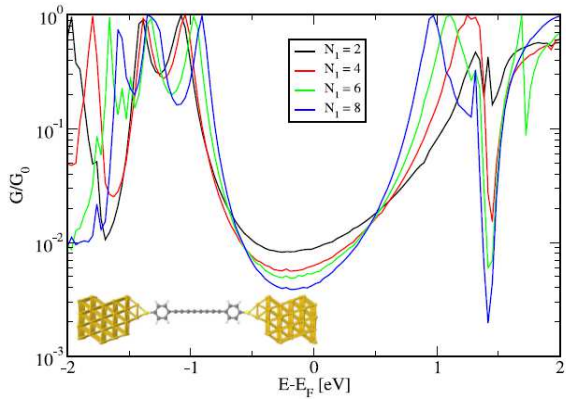}
\caption{\label{olisim}Transmission curves from the
SMEAGOL simulations of oligoynes of varying lengths}
\end{figure}

To demonstrate that this effect is present in atomistic
calculations of electron transport, we compare equation
(\ref{bridgeeven}) with a calculation based on the
 ab-initio transport code SMEAGOL.
This code uses a combination of density functional theory (DFT)
\cite{DFT} and the non-equilibrium Green's function formalism
\cite{NEGF} to calculate the transport characteristics of atomic
scale devices. The DFT Hamiltonian is obtained from the SIESTA
code \cite{Siesta} and is used by SMEAGOL to calculate the
electronic density and the transmission. Within the NEGF the
system is divided in three parts, the left lead, the right lead
and the extended molecule (EM). The EM contains the molecule plus
some layers of gold, whose electronic structure is modified due to
the presence of the molecule and the surfaces and differs from the
bulk electronic structure. The molecular structure consists of an
oligoynes capped with phenyl rings and attached to the electrodes
by thiolate groups. The SMEAGOL results are shown in
Fig.(\ref{olisim}), which clearly possesses a critical energy $E_c
\approx 0.5 {\rm eV}$ at which all curves (at least for the longer
chains) intersect. The analytic expression assumes that the
parameters $\varepsilon_1$ and $\gamma$ describing the chains are
independent of length. In fact the self-consistent DFT parameters
of the shortest chain $(N_1=2)$ differ slightly from those of the
longer chains and therefore the black curve of figure 5 does not
quite pass though the intersection point at $E=E_c$.

Clearly the length independence of even and odd chains leads to an
even-odd oscillation in the electrical conductance of oligoynes,
when $E_f$ is close to $E_c$. This effect has also been observed
in experiments on atomic wires of Au, Pt, and Ir \cite{4}, which exhibit electrical conductance oscillations as a function
of the wire length and similar oscillations as a function
of bias voltage and electrode separation \cite{5,6}. Several theoretical papers \cite{8}-\cite{24}
have also addressed these osillations. The
above analysis also demonstrates that this effect is present in
multi-branch structures, provided the band centres of different
branches occur at the same energy.

\section{Quantum interference in a two-branch molecule}

We now turn to the quantum interference effect transistor (QuIET)
discussed in \cite{QuIET}, which corresponds to the choice $M=2$.
To demonstrate that equation (\ref{eq:greenT}) (or equivalently
(\ref{eq:TE})) reproduces the key features of a QuIET, we compare
it with the results of a detailed simulation using SMEAGOL
\cite{Smeagol}.

\begin{figure}
\includegraphics[width=0.9\columnwidth]{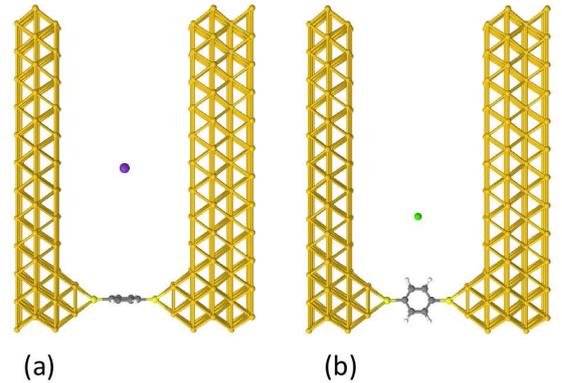}
\caption{\label{struc} Molecular structure used in the transport
simulations with the potassium point charge in configuration C1
(a) and the calcium point charge in configuration C2 (b).}
\end{figure}

\begin{figure}
\includegraphics[width=0.9\columnwidth,scale=0.9]{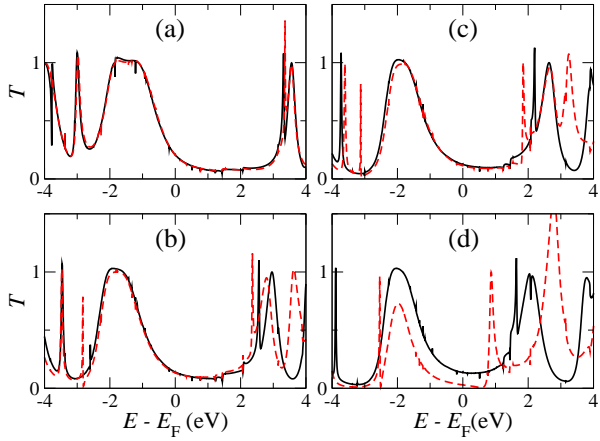}
\caption{\label{K} Transmission functions obtained from SMEAGOL
simulations with the potassium point charge moving closer to the
molecule through (a)-(d). The continuous curve corresponds to C1
and the dashed curve to C2.
}
\end{figure}

The atomic arrangements for the SIESTA/SMEAGOL calculations are shown in Fig. (\ref{struc}). The first arrangement (C1) corresponds to the point
charge located along a line perpendicular to the plane of the
molecule and which passes through its center. In this
configuration  the point charge produces a symmetric voltage which affects the two branches to the same extent. The second arrangement (C2) corresponds to
the point charge located in the plane of the molecule, closer to
one branch of the BDT. In this case the two branches are subject to different electrostatic
potentials, which induces quantum interferences in the electron transmission through the molecule. Both
configurations were simulated using a point charge of either
potassium (K) or calcium (Ca), giving a total of four cases. K and
Ca are alkali and alkaline-earth atoms with 1 and 2 valence
electrons in the last shell, respectively. Due to their high
electropositivity both atoms lose their valence electrons when
they are inserted in the unit cell and become ionized with a
charge of +e and +2e, respectively. The complete removal of the
valence electrons from these atoms can be ensured by reducing the
cutoff radii of their orbitals to 3.5 Bohr, which confine the
electrons in the atom more closely and therefore increase their
energy, making sure they move to lower energy states in the
extended molecule. The basis sets used in the simulation were
single-zeta (SZ) for the point charge and double-zeta polarized
(DZP) for all other elements. The exchange and correlation
potential was calculated with the generalized gradient
approximation (GGA) and the Perdew-Burke-Ernzerhof parametrization
\cite{PBE}. The gold leads were grown along the (001) direction,
and each side of the extended molecule had 3 and 5 layers,
respectively, with 36 atoms (12$\times$3 atoms) per layer. The
molecule was contacted in a hollow configuration to four
additional gold atoms on each side. Since the system was much
larger in the $x$ (3 atoms) than in the $y$ direction (12 atoms to
leave space for the charge to move), 1 $k$-point was used along
$x$ and 4 $k$-points along $y$.

\begin{figure}
\includegraphics[width=0.9\columnwidth,scale=0.9]{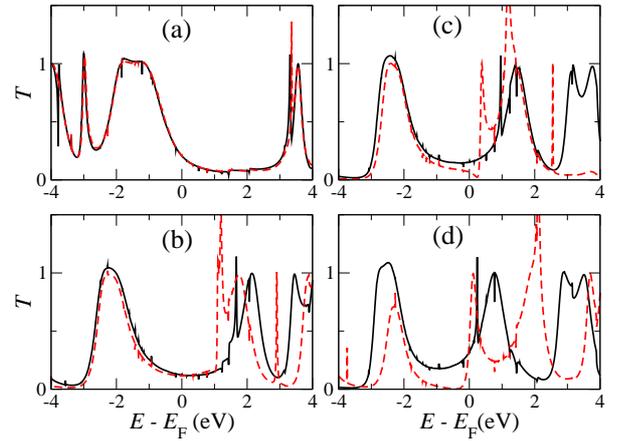}
\caption{\label{Ca} Transmission functions obtained from SMEAGOL
simulations with the calcium point charge moving closer to the
molecule through (a)-(d). The continuous curve corresponds to C1
and the dashed curve to C2.}
\end{figure}

The results are shown in Fig. (\ref{K}) and Fig. (\ref{Ca}),
for potassium and calcium, respectively. Each graph contains two
curves, corresponding to the cases C1 and C2. In plot (a) the
charge is located at a far distance, $\sim 15$ {\AA} from the
molecule, and therefore both C1 and C2 produce the same curve.
From (b) to (d) the charge is gradually moved towards the molecule
(6.29, 5.29 and 4.79 {\AA} away from the center of the ring in
(b), (c) and (d), respectively).

We observe that when the charge moves towards the molecule the
peaks shift in energy in the negative direction due to the
positive potential. However the effect is different depending on
where the charge is located relative to the ring. As can be seen,
there is a clear difference in both Fig. (\ref{K}) and Fig.
(\ref{Ca}) between the continuous and dashed transmission curves
in graphs (b)-(d). An extra peak in the dashed transmission curve
(C2) appears and the height of the HOMO peak is reduced, whereas
the continuous transmission curve (C1) is simply shifted to lower
energies without much change in the resonances. Also, through
comparison of (a) and (d) we notice a clear narrowing of the HOMO
and broadening of the LUMO peak in all cases. We observe a clear reduction
of the transmission at the Fermi energy when the charge is located closer to
one arm of the molecule (C2). In contrast, for system C1, there is very little
change of the  transmission about the Fermi energy, because the point charge
produces the same phase shifts in the two branches and therefore does not
modify interference effects associated with coherent superposition of waves
propagating along separate paths.

We also checked the projected density of states (PDOS) on each
branch of the BDT to see the specific effect of the charge on the
electronic structure in each case. In C1 the PDOS on each branch
remains equally distributed and simply shifts to lower energies.
However, in C2 there is a clear difference in the PDOS on each
branch; the PDOS on the closest branch to the charge is more
affected and shifted to lower energies than the PDOS of the
opposite branch. This supports the observation of the previously
suggested QIE.

\begin{figure}
\includegraphics[width=0.9\columnwidth]{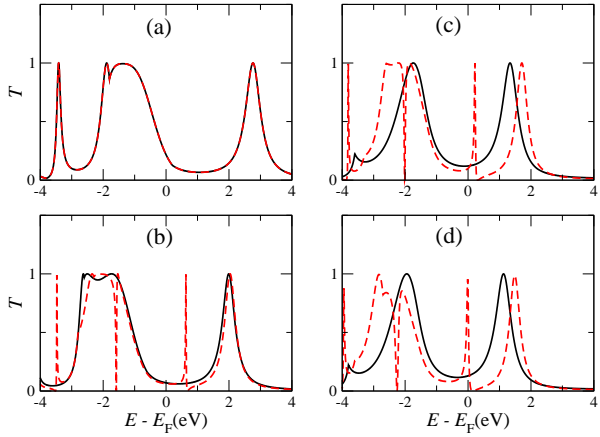}
\caption{\label{tbm} Transmission curves for the tight binding
model. The continuous line corresponds to the case where the
charge is located along a line perpendicular to the ring and which
passes through its center. The dashed line corresponds to the case
where the charge is closer to one arm. Panel (a) shows the
transmission curves when the charge is away from the molecule.
From panel (b) to (d) the onsite energy parameters are changed to
simulate the charge moving towards the molecule.}
\end{figure}

To elucidate the underlying physics, we employ Eq.
(\ref{eq:greenT}) to model electron transmission through a
two-branch structure. In the absence any charge, we choose the
hopping parameters $\alpha_L = \beta_R = 1.5$, $\gamma_L=\gamma_R
= 4.0$ and $\gamma_1 = \gamma_2 = 1.0$ and  the on-site energies
$\varepsilon_L^0 = \varepsilon_R^0 = 2.0$, $\varepsilon_1 = 0.0,
\varepsilon_2 = 0.0$ and $\varepsilon_L = \varepsilon_R = 4.0$.
This leads to the transmission curve shown in Fig. (\ref{tbm})
(a), which is very close to the ab-initio result.  In
configuration C1, where a charge affects both branches equally,
the presence of a charge is modelled by shifting the on-site
energies $\varepsilon_L$, $\varepsilon_1$, $\varepsilon_2$ and
$\varepsilon_R$, upwards or downwards by the same amount, which
depends on the sign and strength of the charge. The outcome
produced by a positive charge is represented by the continuous
transmission curves in Fig. (\ref{tbm}). The charge moves closer
to the ring from (b) to (d) and the parameters are chosen as
follows (b) $\varepsilon_L^0 = \varepsilon_R^0 = 1.4$,
$\varepsilon_1 = \varepsilon_2 = -0.8$, (c) $\varepsilon_L^0 =
\varepsilon_R^0 = 1.2$, $\varepsilon_1 = \varepsilon_2 = -1.8$,
(d) $\varepsilon_L^0 = \varepsilon_R^0 = 1.0$, $\varepsilon_1 =
\varepsilon_2 = -2.0$. In each of these plots, $\varepsilon_L =
\varepsilon_R$ remain unchanged throughout. As in the ab-initio
simulations we see that the entire transmission curve is shifted
to lower energies and quantum interference effects are negligible.
Interestingly, as a consequence of this shift and the
corresponding change in the electronic structure, the width of the
variability in the local density of states at the contact, the
width of the HOMO decreases and the width of the LUMO increases,
in agreement with the ab-initio results.

To produce quantum interference, we now examine the effect of a scanning point charge
 placed in configuration C2; i.e. closer to one branch of the
ring. To model this effect using Eq. (\ref{eq:TE}), the parameters
are now adjusted asymmetrically; i.e. they are changed less in the
branch which is far away from the charge and more in the branch
which is closer. The adjustment also includes changing the contact
points $\varepsilon_L^0$ and $\varepsilon_R^0$ as these will feel
a smaller effect from the charge than the nearer branch. The
adjusted parameters are chosen as follows (b) $\varepsilon_L^0 =
\varepsilon_R^0 = 1.35$, $\varepsilon_1 = -1.0$, $\varepsilon_2 =
-0.5$ (c) $\varepsilon_L^0 = \varepsilon_R^0 = 1.1$,
$\varepsilon_1 = -1.5$, $\varepsilon_2 = -0.8$ (d)
$\varepsilon_L^0 = \varepsilon_R^0 = 1.0$, $\varepsilon_1 = -1.8$
and $\varepsilon_2 = -0.9$. As before, $\varepsilon_L =
\varepsilon_R$ and are unchanged. The transmission corresponding
to these parameters is shown Fig.(\ref{tbm}) (dashed curves),
where the point charge is brought successively closer to the
molecule from (b) to (d). We see again from (a) through to (d)
that the peaks have all shifted to lower energies, but the HOMO
dramatically changes and reduces its height. Also, an additional
peak appears due to the point charge effect on the electronic
structure on only one arm of the molecule, which causes
interferences in the transmission through the system. This again
agrees with the SMEAGOL simulations and suggests this analytical
model captures the qualitative features of transmission in
ring-like molecules.

\begin{figure}
\includegraphics[width=0.9\columnwidth]{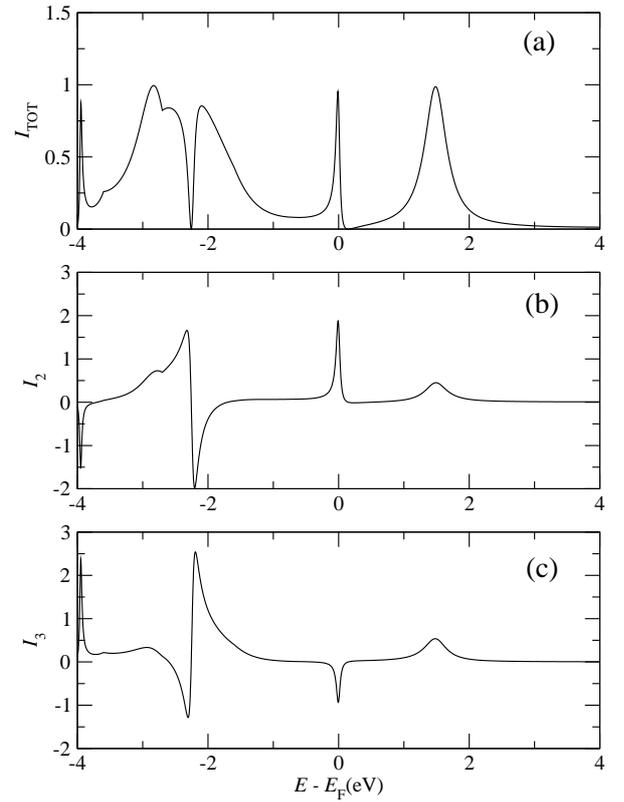}
\caption{\label{I} (a) Total current through the molecule, (b),
and the current through branch 1, (c), and  branch 2 of the
molecule. Parameters used are exactly as the dashed red line in
Fig 9(d), namely $\varepsilon_R = \varepsilon_L = 4.0$,
$\varepsilon_L^0= \varepsilon_R^0 = 1.0$, $\varepsilon_1 = -1.8$,
$\varepsilon_2 = -0.9$ $\gamma_L = \gamma_R = 4.0$, $\gamma_1 =
\gamma_2 = 1.0$, $\alpha_L = \beta_R = 1.5$. }
\end{figure}

Having established that the analytical model captures the
essential features of the ab-initio simulations, we now show how
this model can be employed to examine the internal currents within
different branches of the molecule, which are obtained by
evaluating equation (\ref{eq:il}). When the ion located close to
branch $1$, the lower graphs of Fig. (\ref{I})  show the internal
currents $I_1$ and $I_2$ through the individual branches, whereas
the upper graph shows the total current $I_{\rm TOT}=I_1+I_2 =
T(E)$.
 Figs. (\ref{I}) (b) and (c) clearly demonstrate that the
current in a single branch can greatly exceed the total current through
the molecule when a counter current of opposite sign occurs in the other
branch of the molecule and can clearly exceed the upper bound of $I_{\rm TOT}=1$. The appearance of such unbounded counter currents is yet another manifestation of quantum interference within single molecules.
\cite{loop}.

\section{Summary}

In conclusion, we have presented ab initio simulations and an
 analytical formula, which highlights a range of
interference effects in single and multi-branch structures. The
analytical solution is rather versatile and has the advantage that
it can be evaluated on a pocket calculator. It provides insight
into length-independent electrical conductances for even and odd
oligoyne chains, when the Fermi energy coincides with the band
centre of the oligoyne bridge and allows us to predict that this
behaviour is also present in multi-branch structures, provided the
branches share a common band centre.
As demonstrated in the manuscript the energy $E_c$ at which this odd-even effect occurs corresponds to $k1= \pi/2$. This condition is very general and is independent of the nature of the orbitals. For the particular case of oligoynes,  this is a consequence of $\pi$ coherent transport, but for other systems, such a metallic wires, this would not be the case. The case $M=1$ demonstrates that quantum interference does not require the presence of physically different paths, because even in this case, interference due to scattering from nodal impurity sites and connections to external leads arises from the amplitudes $a_L$ and $a_R$ in equations (14) and (15). Both the magnitudes and phases of these amplitudes appear on the right hand side of equation (7) and therefore even for a single-branch system, they contribute to interference.

Ab initio simulations based
on density functional theory, demonstrate the presence of quantum
interference in BDT, due to electrostatic interactions associated
with a scanning point charge positioned close to the molecule. We
have shown that a scanning charge located within the plane of a
BDT molecule produces a sizeable quantum interference, whereas a
charge approaching the molecule along a line perpendicular to the
plane produces a much smaller effect, in agreement with the
analytical formula. In spite of the consistency between the TB result and the ab initio result for the BDT system, there are of course quantitative differences between them. In part this arises because the tight-binding model includes only a single ("$\pi$") orbital per atom, whereas the ab initio description includes both $\pi$ transport and $\sigma$ tunneling. In addition, the tight-binding model includes only a single scattering channel in each lead, whereas the ab initio model contains multiple channels.

Using the analytical model, we have also
investigated the internal currents within a two-branch molecule
and demonstrated that large currents and counter currents can
occur, which exceed an upper bound for the total current through
the molecule.

\acknowledgments We wish to thank the Spanish Ministerio de Ciencia e Innovaci\'on, the UK EPSRC and the European Research Networks NanoCTM and FUNMOLS for funding.

\section{Appendix I: Derivation of equation \ref{eq:greenT} for transmission though the multi-branch structure of figure (1).}

We derive the  equation for $T(E)$ by matching wave functions at
the nodes of a multi-branch structures and later make a comparison
with results obtained from a corresponding Green's function
analysis. The starting point is the tight binding Schr\"{o}dinger
equation, which can be written

\begin{equation}\label{eq:cryst}
\varepsilon_\mu \psi_\mu - \sum_\nu\gamma_{\nu,\mu} \psi_{\nu}
 = E \psi_\mu,
\end{equation}

\noindent where the summation is over  all nearest neighbours
$\nu$ of site $\mu$. Choosing $\mu$ to label the site just to the
left of the nodal site $L$ (whose wave function is denoted
$\chi_L$) yields $\varepsilon_L \psi_{-1}^{(L)} - \gamma_L
\psi_{-2}^{(L)} - \alpha_L\chi_L = E \psi_{-1}^{(L)}$, where
$\gamma_L =\gamma{-1,-2}$ and $\alpha_L=\gamma_{L,-1}$. From this
expression, and noting that the Schr\"{o}dinger equation in the
left lead takes the form of a recurrence relation \cite{recur},
the wave function at the node $L$ is given by

\begin{equation}\label{eq:chi}
\chi_L =\frac{\gamma_L(1+r)}{\alpha_L}
\end{equation}

\noindent Similarly, choosing $\mu$ to label the site just to the
right of the nodal site $R$ (whose wave function is denoted
$\chi_R$) yields

\begin{equation}\label{eq:theta}
\chi_R = \frac{\gamma_R t}{\beta_R}.
\end{equation}

\noindent Choosing $\mu$ to label the first site ($1_l$) of chain
$l$ yields for all $l$,
\begin{equation}\label{eq:chi1}
\chi_L =(\gamma_l/\alpha_l)(A_l+B_l)
\end{equation}

\noindent and choosing $\mu$ to label the last site  ($N_l$ of
chain $l$ yields for all $l$,
\begin{equation}\label{eq:theta1}
\chi_R =(\gamma_l/\beta_l)(A_le^{ik_l(N_l+1)}+B_le^{-ik_l(N_l+1)})
\end{equation}

\noindent Finally choosing $\mu$ to label the nodal sites  $L$ and
$R$ yields
\begin{equation}\label{eq:chiL2}
\varepsilon^0_L\chi_L - \sum_{l=1}^M \alpha_l\psi^{(l)}_1
-\alpha_L( e^{-ik_L}+ r(E)e^{ik_L})=E\chi_L
\end{equation}
\noindent and

\begin{equation}\label{eq:chiR2}
\varepsilon^0_R\chi_R - \sum_{l=1}^M \beta_l\psi^{(l)}_{N_l}
-\beta_R t(E)e^{ik_R}=E\chi_R
\end{equation}

\noindent Equations (\ref{eq:chi}) with (\ref{eq:chiL2}) and(\ref{eq:theta}) with
(\ref{eq:chiR2}) combine to yield

\begin{equation}\label{eq:a}
\chi_L=\frac{\gamma_L}{\alpha_L}(1+r) =
\frac{\sum_{l=1}^M\alpha_l\psi_1^{(l)}  - 2\alpha_L i\sin
(k_L)}{a_L}
\end{equation}

\noindent and

\begin{equation}\label{eq:b}
\chi_R=\frac{\gamma_R}{\beta_R} t =
\frac{\sum_{l=1}^M\beta_l\psi_{N_l}^{(l)}}{a_R}
\end{equation}

\noindent where $a_L$ and $a_R$ are given by equations
(\ref{eq:aL}) and (\ref{eq:aR}).

From the form of the wave functions in the branches, given by
equation (2), these can be written

\begin{eqnarray}\label{eq:matrix}
 \left( \begin{array}{c}
\chi_L\ \\
\chi_R
\end{array}\right)&=
&\frac{1}{a_L} \left( \begin{array}{c}
-2i\alpha_L\sin (k_L) \\
0 \end{array} \right)\\
& & + \sum_{l=1}^M\left( \begin{array}{cc}
\frac{\alpha_l e^{ik_l}}{a_L}  & \frac{\alpha_l e^{-ik_l}}{a_L} \\
\frac{\beta_l e^{ik_lN_l}}{a_R} & \frac{\beta_l e^{-ik_lN_l}}{a_R} \end{array}
\right) \left( \begin{array}{c}
A_l \\
B_l \end{array} \right)\nonumber
\end{eqnarray}

\noindent Since equations (\ref{eq:chi1}) and (\ref{eq:theta1}) yield

\begin{equation}\label{eq:matrix22}
\left( \begin{array}{c}
A_l\\
B_l
\end{array}\right)=
\frac{1}{-2i\gamma_l \sin k_l(N_l+1)} \left( \begin{array}{cc}
\alpha_l e^{-ik_l(N_l+1)}  & -\beta_l\\
-\alpha_le^{ik_l(N_l+1)}& \beta_l \end{array} \right)\left(
\begin{array}{c}
\chi_L \\
\chi_R
\end{array}\right)
\end{equation}

\noindent $A_l$ and $B_l$ can be eliminated from equation (\ref{eq:matrix}) to yield

\begin{equation}\label{eq:matrix2}
\\
\left( \begin{array}{c}
\chi_L\ \\
\chi_R
\end{array}\right)=
G
\left( \begin{array}{c}
2i\alpha_L\sin (k_L) \\
0 \end{array} \right) =
 \left( \begin{array}{c}
 G_{LL}\\
G_{RL}\end{array} \right)2i\alpha_L\sin (k_L)
\end{equation}

\noindent In this expression, the matrix $G$ has the form

\begin{eqnarray}\label{eq:G}
G= \left( \begin{array}{cc}
G_{LL}  & G_{LR}\\
G_{RL}& G_{RR} \end{array}
\right)
\end{eqnarray}

\noindent and is given by

\begin{eqnarray}\label{eq:G11}
G^{-1}=
 \left( \begin{array}{cc}
{-a_L}  & 0 \\
 0& {-a_R} \end{array}
\right) +\left( \begin{array}{cc}
{x_L}  & y \\
 y& {x_R} \end{array}
\right)
  \nonumber\\
\end{eqnarray}

\noindent where $x_L$, $x_R$ and $y$ are given by equations
(\ref{eq:greeny}), (\ref{eq:greenxL}) and (\ref{eq:greenxR}). From
this expression, one obtains $\chi_R$ and hence the transmission
amplitude $t$, via equation (\ref{eq:theta}).

The physical meaning of the various contributions to the above
expressions can be understood by carrying out a parallel analysis
based on Green's functions \cite{andres,brand}, which reveals that
equation (\ref{eq:G11}) is simply Dyson's equation for the Green's
function matrix elements involving the nodal sites $L$ and $R$. Comparison with Refs. \cite{andres,brand} also demonstrates that $v_L (\alpha_L/\gamma_L)^2$ and $v_R(\alpha_R/\gamma_R)^2$ in Eq,(5) are imaginary parts of the self-energies of the left- and right-hand electrodes, respectively.

This is demonstrated by noting that the Green's function for a
finite linear chain of $N_l$ sites, with nearest-neighbour hopping
elements $-\gamma_l$ and diagonal elements $\varepsilon_l$ is

\begin{equation}\label{eq:greenld}
g_l(n_l,n'_l)=  \begin{array}{cc}
A_l \sin k_ln_l \sin k_l[n'_l-(N_l+1)]  & ({\rm for}\, n_l\le n'_l)\\
A_l \sin k_ln'_l \sin k_l([n_l-(N_l+1)]& ({\rm for}\, n'_l\le n_l)\\
\end{array}
\end{equation}

\noindent where $A_l=1/[\gamma_l\sin k_l\sin k_l(N_l+1)]$.
An alternative form of this expression is
$g_l(n_l,n'_l)=(A_l/2)(\cos k_l[N_l+1-|n_l-n'_l|] - \cos k_l[N_l+1-n_l-n'_l])$

The quantity $g_l(n_l,n'_l)$ is the Greens function
matrix element connecting atom $n_l$ to atom $n'_l$ of the
decoupled branch $l$, which would arise when $\alpha_l=\beta_l=0$.
The off-diagonal matrix element describing propagation from one
end of such a branch to the other is

\begin{equation}\label{eq:greenylg}
g_l(1_l,N_l)= -\sin k_l/{\gamma_l\sin k_l (N_l+1)}
\end{equation}

\noindent whereas the diagonal matrix element evaluated on an end
atom is

\begin{equation}\label{eq:greenxl1R}
g_l(1_l,1_l)=g_l(N_l,N_l) = -\sin k_l N_l/{\gamma_l\sin k_l
(N_l+1)}
\end{equation}

\noindent As expected, these quantities diverge when $\sin k_l
(N_l+1)=0$, which corresponds to the eigenenergies of an isolated
branch. In terms of these Greens
functions,

\begin{equation}\label{eq:greenyl1} y_l =
-\alpha_l\beta_l g_l(1,N_l)
\end{equation}

\begin{equation}\label{eq:greenxlL1}
x^L_l =  -\alpha_l^2 g_l(1_l,1_l)
\end{equation}

\noindent and

\begin{equation}\label{eq:greenxlR1}
x^R_l =  -\beta_l^2 g_l(N_l,N_l)
\end{equation}

Within a Green's function approach, one defines the nodal self energy
matrix $\sigma$ to be

\begin{equation}\label{eq:sigma}
\sigma = \sum_{l=1}^M \sigma_l
\end{equation}

\noindent where $\sigma_1$ is the contribution to the self-energy
from branch $l$, given by

\begin{eqnarray}\label{eq:sigmal}
\sigma_l=
 \left( \begin{array}{cc}
{-\alpha_l}  & 0 \\
 0& {-\beta_l} \end{array}
\right)g_l\left( \begin{array}{cc}
{-\alpha_l}  & 0 \\
 0& {-\beta_l} \end{array}
\right)
  \nonumber\\
\end{eqnarray}

\noindent In this expression $g_l$ is the Greens function
connected the end atoms of an isolated branch:

\begin{equation}\label{eq:gl}
g_l=
 \left( \begin{array}{cc}
g_l(1_l,1_l)  & g_l(1_l,N_l) \\
g_l(N_l,1_l)& g_l(N_l,N_l) \end{array} \right)
\end{equation}

\noindent This demonstrates that

\begin{equation}\label{eq:sigma2}
\left( \begin{array}{cc}
x_L  & y \\
y& x_R \end{array} \right)=-\sigma
\end{equation}

\noindent and therefore equation (\ref{eq:G11}) takes the form of
Dyson's equation:

\begin{equation}\label{eq:G222}
G^{-1}=
 \left( \begin{array}{cc}
{g^{-1}_L}  & 0 \\
 0& {g^{-1}_R} \end{array}
\right) -\sigma
\end{equation}

\noindent where $g_L  = -a^{-1}_L$ and $g_R=-a_R^{-1}$ are
diagonal elements of the Greens function of the decoupled
semi-infinite chains (obtained by setting all
$\alpha_l=\beta_l=0$), evaluated on the left (L) and right (R)
nodal sites respectively. This also demonstrates that the form of equation(\ref{eq:greenT}) and in particular $G_{RL}$ does not change even when the branches $l$ are replaced by arbitrary elastic scattering regions, connected to nodal sites by bonds $\alpha_l$ and $\beta_l$, provided $g_l$ is replaced by the Green's function of the $l$th scattering region. With this redefinition of $y_l$, the condition for destructive interference ($y=0$) remains unchanged. For example, if instead of a linear chain of $n_l$ sites, branch $l$ is replaced by a loop of $n_l$ sites, , then
equation (\ref{eq:greenld}) is replaced by the Greens function of a linear chain of $n_l$ sites with periodic boundary conditions, namely
$g_l(n_l,n'_l)=(\cos k_l[N_l/2-|n_l-n'_l|])/(2\gamma \sin k_l \sin [k_lN_l/2])$ and equation (\ref{eq:gl}) is replaced by

\begin{equation}\label{eq:gloop}
g_l=
 \left( \begin{array}{cc}
g_l(n_l,n_l)  & g_l(n_l,m_l) \\
g_l(m_l,n_l)& g_l(m_l,m_l) \end{array} \right),
\end{equation}
where $n_l$ and $m_l$ label the sites of the loop connected to the nodal sites $L$ and $R$ respectively. Taking this to an extreme, any of the branches $l$ could even be replaced by a multi-branch scatterer, simply by replacing $g_l$ by the Greens function of an isolated multi-branch system, obtained from $G$ by setting $\alpha_L=\beta_R=0$.

The above analysis, which focusses on the wave-like nature of Greens functions is rather different in spirit from alternative approaches which emphasise the algebraic nature of Greens functions, which for finite structures, take the form of ratios of polynomials, whose denominator is proportional to the secular equation \cite{ratner}. To make contact with this approach, we note that equation (\ref{eq:G222}) yields

\begin{equation}\label{eq:GG22}
G=
 \frac{-1}{\Delta}\left( \begin{array}{cc}
x-a_R  & -y\\
-y& x-a_L \end{array} \right)
\end{equation}

\noindent where $\Delta=\Delta_1 +i\Delta_2$, and therefore
 the equation
$\Delta_1=0$ is the secular equation for the isolated multi-branch
structure, which arises when $\alpha_L=\beta_R=0$. More generally,
from equations (\ref{eq:dl}) and (\ref{eq:d2}),
 the equation $\Delta_1 =\tilde\Gamma_L\tilde\Gamma_R$ is
the secular equation for the same isolated system, but with the
site energies of the nodal atoms shifted by the real part of their
respective self energies.

Finally the current per unit energy in branch $l$, carried by
electrons of energy $E$ injected from the left lead is
$(2e/h)I_l$, where

\begin{equation}\label{eq:III1}
I_l = \frac{v_l}{v_L}(|A_l|^2 - |B_l|^2)
\end{equation}

\noindent Expressions for $A_l$ and $B_l$ are obtained from equation
(\ref{eq:matrix22}), which combine to yield equation(\ref{eq:il})
of the main text.

The above comparison between the wave-function-matching and
Green's function underpins a deep understanding of equation
(\ref{eq:matrix2}), because if $\mu \le -1$ labels a site in the
left lead and $\nu \ge \mu$ labels a site inside the scattering
region or in the right lead, then the wave function $\psi_\nu$ is
related to $G_{\nu,\mu}$ by the expression

\begin{equation}\label{gpsi}
\psi_\nu=2i\gamma_L \sin k_L e^{ik_L \mu}G_{\nu,\mu}
\end{equation}
Furthermore, starting from the limit $\alpha_L=0$ and then
including the effect of $\alpha_L$ via Dyson's equation yields
$G_{LL}=(\gamma_L/\alpha_L)e^{-ik_L}G_{L,-1}$ and
$G_{RL}=(\gamma_L/\alpha_L)e^{-ik_L}G_{R,-1}$. Hence equation
(\ref{eq:matrix2}) can be written in the intuitive form

\begin{equation}
\left( \begin{array}{c}
\chi_L\\
\chi_R\end{array} \right)
=
\left( \begin{array}{c}
G_{L,-1}\\
G_{R,-1}\end{array} \right)e^{-ik_L}2i\gamma_L\sin (k_L),
\end{equation}
\noindent which is simply an example of equation (\ref{gpsi}),
with $\mu=-1$ and $\nu = L$ or $R$.

 As mentioned in the main text,
equation(\ref{eq:greenT}) is extremely versatile.
For example, the case of $M=1$ $N_1 > 1$, can  be used to describe a donor-bridge-acceptor molecules.
In this case, to obtain a simple description of rectification, all parameters should be assigned and
appropriate dependence on the applied voltage $V$. The simplest
model is obtained by setting $\epsilon_L(V)=\epsilon_L(0) +eV/2$,
 $\epsilon^0_L(V)=\epsilon^0_L(0) +eV/2$,
 $\epsilon_R(V)=\epsilon_R(0) -eV/2$
 $\epsilon^0_R(V)=\epsilon^0_R(0) -eV/2$, and then computing the
 current via the expression $I=\int^{E_F+eV/2}_{E_F-eV/2}T(E) dE$.

\begin{figure}
\includegraphics[width=0.9\columnwidth]{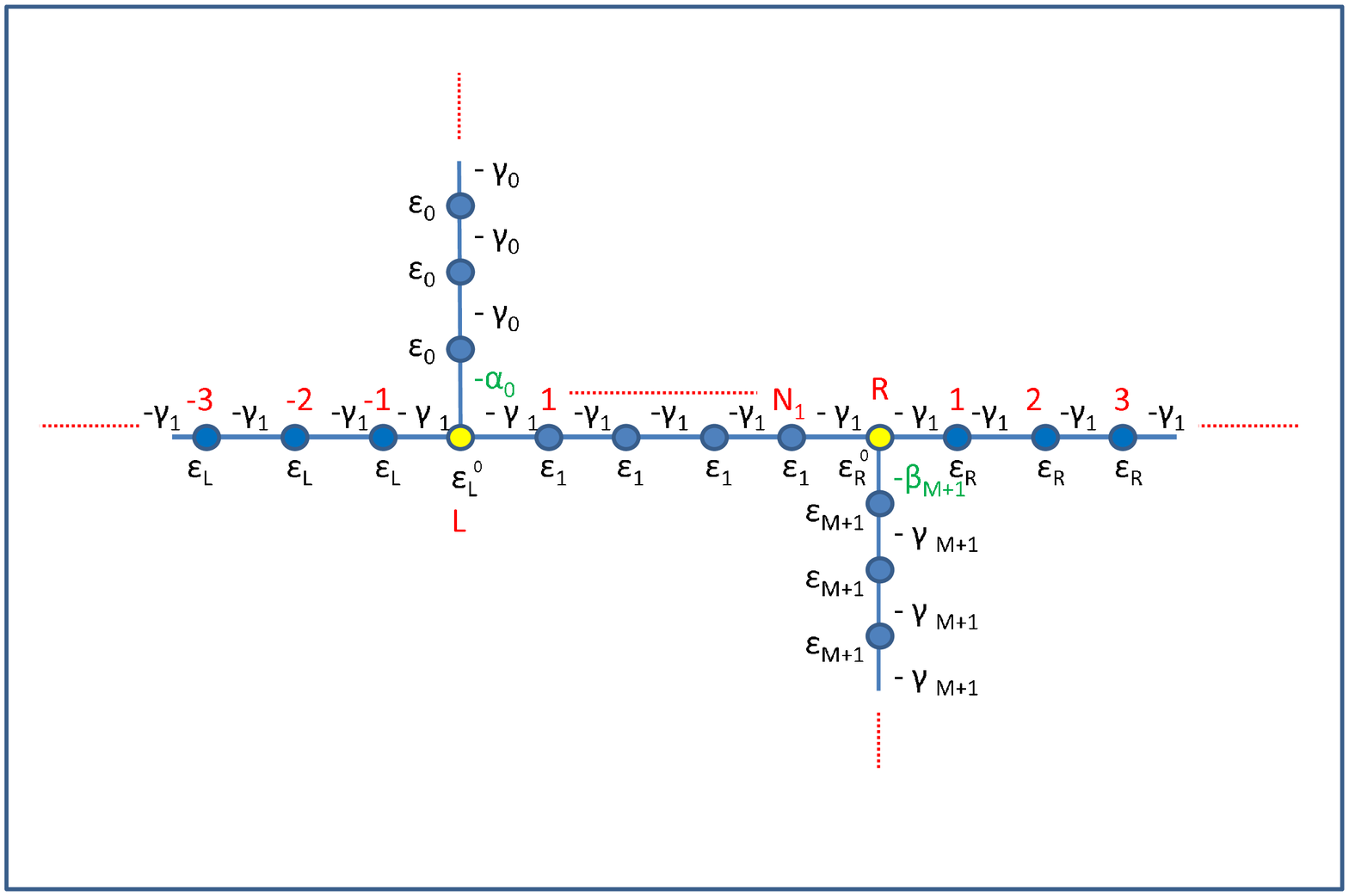}
\includegraphics[width=0.9\columnwidth]{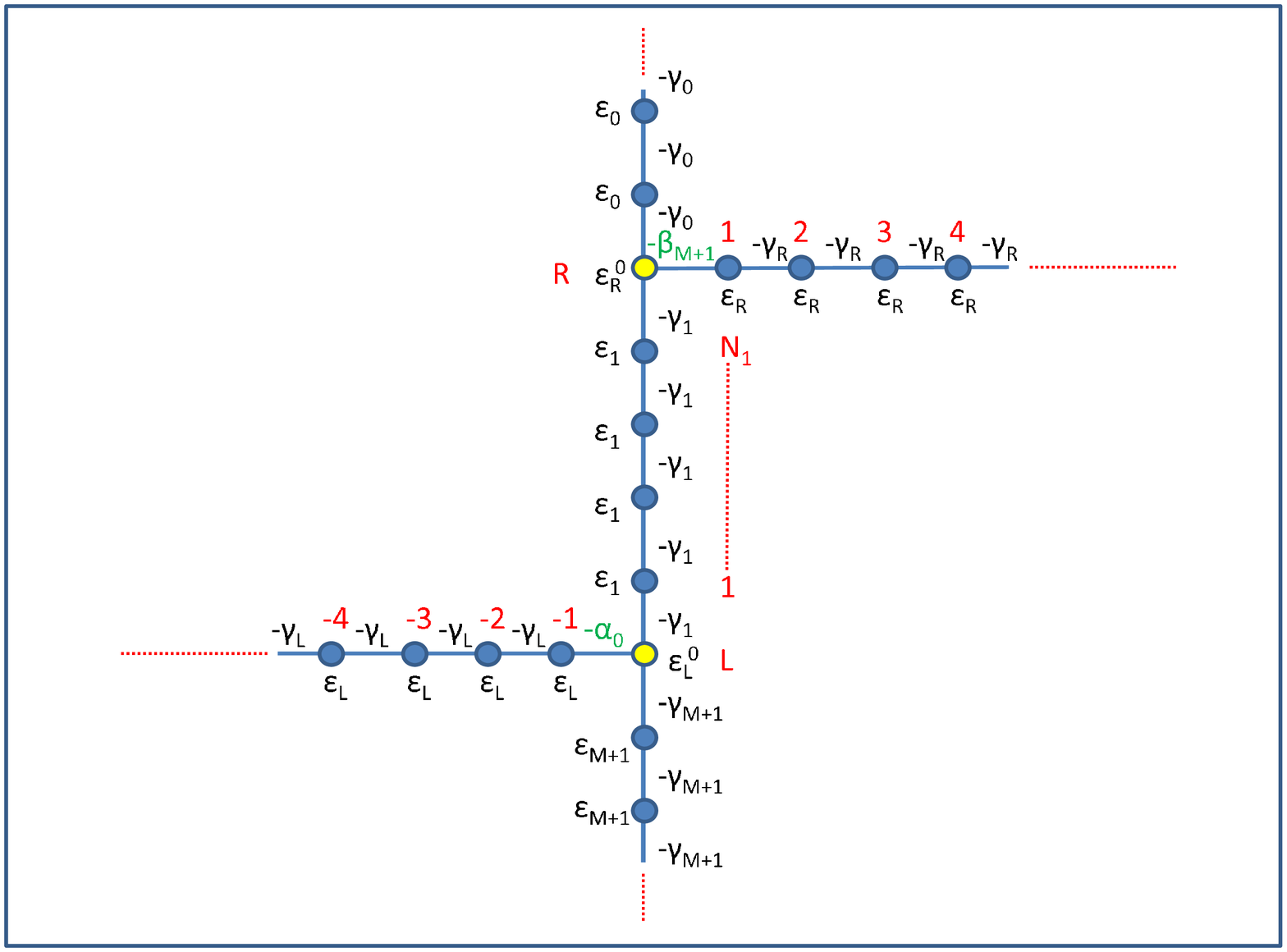}
\caption{\label{Ex3}A diagram to show the case of a molecule with
dangling branches. The top and bottom panels show two equivalent
representations of the same system.}
\end{figure}

To further demonstrate the versatility of equation(\ref{eq:il}), we end this appendix by noting that it readily describes the effect of Fano resonances on transport and the effect of coupling to a molecule at different points along its length. To illustrate this, consider a
structure in which dangling  branches,
labelled $l=0$ and $l=M+1$, are attached by couplings $\alpha_0$ and
$\beta_{M+1}$ to the nodal sites on the left and right
respectively, as shown in figure (\ref{Ex3}).

In this case, equations (\ref{eq:greenT}) and (\ref{eq:TE}) are
unchanged, except that $\varepsilon_R^0$ and $\varepsilon_L^0$ are
renormalised by the self energies of the dangling branches, and
replaced by

\begin{equation}\label{epsiR}
\tilde\varepsilon_R^0=\varepsilon_R^0 +\beta_{M+1}^2 g_{M+1}
\end{equation}

\noindent and

\begin{equation}\label{epsiL}
\tilde\varepsilon_L^0=\varepsilon_L^0 +\alpha_0^2 g_0,
\end{equation}

\noindent where

\begin{equation}\label{gL}
g_0=-\frac{\sin k_0N_0}{\gamma_0 \sin k_0(N_0+1)}
\end{equation}

\noindent and

\begin{equation}\label{gR}
g_{M+1}=-\frac{\sin k_{M+1}N_{M+1}}{\gamma_{M+1} \sin
k_{M+1}(N_{M+1}+1)}
\end{equation}

\noindent Clearly an anti-resonance occurs when the energy $E$
coincides with an eigen-energy of either of the two branches,
because at these energies, one of the Green's functions $g_0$ or
$g_{M+1}$ diverges and therefore one of the renormalised nodal
site energies $\tilde\varepsilon_R^0$ or $\tilde\varepsilon_L^0$
diverges. This is equivalent to introducing an infinite potential
at one of the nodes and therefore at these energies, $T(E)$
vanishes. This behaviour arises from the interaction between bound
states in the dangling branches and the continuum of states
associated with the external leads and is typical of a Fano
resonance.

By redrawing Figure figure (\ref{Ex3}a) as shown in figure
(\ref{Ex3}b), one can see that the above equation describes a
linear molecule contacted at atoms within the length of the
molecule, rather than simply at the end atoms. As an example,
consider the case when $M=1$,
$\varepsilon_1=\varepsilon_0=\varepsilon_{2}=\varepsilon_L=\varepsilon_R$,
$\gamma_1=\gamma_0=\gamma_{2}=\alpha_0=\beta_{2}=\alpha_1=\beta_1$.
The system then comprise a linear chain of length $L=N_0+N_1+N_2
+2$ sites, connect to external leads by nodal sites located at
positions $N_0+1$ and $N_1+N_0+2$ along the chain. By varying
$N_0$ and $N_1$ but with fixed $L$, the expression for $T(E)$ then
describes quantum interference effects which arise when external
leads are connected to a fixed length molecule, at different
locations along its length.

\end{document}